\documentclass[aps,twocolumn,superscriptaddress,floatfix,pre]{revtex4}

\usepackage{amsmath,amssymb,graphicx}
\usepackage{algorithmic}
\usepackage{enumerate}

\newcommand{\pa}{PA}
\newcommand{\sa}{SA}
\newcommand{\pt}{PT}
\newcommand{\GS}{0}
\newcommand{\fGS}{g}
\newcommand{\etaGS}{\eta}
\newcommand{\pGS}{\mathcal{P}}
\newcommand{\eGS}{E_{\GS}}

\begin{document}

\title{Comparing Monte Carlo methods for finding ground states of Ising
spin glasses: \\ Population annealing, simulated annealing, and parallel
tempering}

\author{Wenlong Wang}
\email{wenlong@physics.umass.edu}
\affiliation{Department of Physics, University of Massachusetts,
Amherst, Massachusetts 01003 USA}

\author{Jonathan Machta}
\email{machta@physics.umass.edu}
\affiliation{Department of Physics, University of Massachusetts,
Amherst, Massachusetts 01003 USA}
\affiliation{Santa Fe Institute, 1399 Hyde Park Road, Santa Fe, New
Mexico 87501 USA}

\author{Helmut G. Katzgraber}

\affiliation{Department of Physics and Astronomy, Texas A\&M University,
College Station, Texas 77843-4242, USA}
\affiliation{Materials Science and Engineering, Texas A\&M University, 
College Station, Texas 77843, USA}
\affiliation{Santa Fe Institute, 1399 Hyde Park Road, Santa Fe, New
Mexico 87501 USA}

\begin{abstract}

Population annealing is a Monte Carlo algorithm that marries features
from simulated-annealing and parallel-tempering Monte Carlo. As such, it
is ideal to overcome large energy barriers in the free-energy landscape
while minimizing a Hamiltonian. Thus, population-annealing Monte Carlo
can be used as a heuristic to solve combinatorial optimization problems.
We illustrate the capabilities of population-annealing Monte Carlo by
computing ground states of the three-dimensional Ising spin glass with
Gaussian disorder, while comparing to simulated-annealing and
parallel-tempering Monte Carlo. Our results suggest that population
annealing Monte Carlo is significantly more efficient than simulated
annealing but comparable to parallel-tempering Monte Carlo for finding
spin-glass ground states.

\end{abstract}

\maketitle

\section{Introduction}

Spin glasses present one of the most difficult challenges in statistical
physics \cite{book}. Finding spin-glass ground states is important in
statistical physics because some properties of the low-temperature
spin-glass phase can be understood by studying ground states. For
example, ground-state energies in different boundary conditions have
been used to compute the stiffness exponent of spin glasses
\cite{stiffness1,stiffness2,stiffness3}. More generally, the problem of
finding ground states of Ising spin glasses in three and more dimensions
is a nnondeterministic polynomial-time (NP) hard combinatorial
optimization problem \cite{NPhard} and is thus closely related to other
hard combinatorial optimization problems \cite{MoMe11}, such as protein
folding \cite{protein_folding} or the traveling salesman problem. As
such, developing efficient algorithms to find the ground state of a
spin-glass Hamiltonian---as well as related problems that fall into the
class of ``quadratic unconstrained binary optimization
problems''---represents an important problem across multiple
disciplines.

Many {\em generally applicable} computational methods have been
developed to solve hard combinatorial optimization problems. Exact
algorithms that efficiently explore the tree of system states include
branch-and-cut \cite{Branch_and_cut} algorithms. Heuristic methods
include genetic algorithms \cite{hartmann:01,hartmann:04}, particle
swarm optimization \cite{GA_and_BPSO}, and extremal optimization
\cite{BoPe01,Middleton04}.  The focus of this paper is on heuristic
Monte Carlo methods based on thermal annealing approaches.  In
particular, we studied simulated annealing \cite{SAMC},
parallel-tempering Monte Carlo \cite{ptmc1, ptmc2, ptmc3}, and
population-annealing Monte Carlo \cite{F}. The first two methods are
well-known and have been successfully applied to minimize Hamiltonians,
while the third has been much less widely used in statistical physics
and a primary purpose of this paper is to introduce population annealing
as an effective method for finding ground states of frustrated
disordered spin systems.

Population-annealing Monte Carlo 
\cite{F,A,B,ZhCh10,WaMaKa14} is closely
related to simulated annealing and also shares some similarities with
parallel tempering. Both simulated annealing and population annealing
involve taking the system through an annealing schedule from high to low
temperature. Population annealing makes use of a {\em population} of
replicas of the system that are simultaneously cooled and, at each
temperature step, the population is resampled so it stays close to
the equilibrium Gibbs distribution. The resampling step plays the same
role as the replica exchange step in parallel-tempering Monte Carlo.  On
the other hand, population annealing is an example of a sequential Monte
Carlo algorithm \cite{DoFrGo01}, while parallel tempering is a Markov
chain Monte Carlo algorithm. In a recent large-scale study
\cite{WaMaKa14}, we have thoroughly tested population-annealing Monte
Carlo against the well-established parallel tempering-Monte Carlo
method. Simulations of thousands of instances of the Edwards-Anderson
Ising spin glass show that population annealing Monte Carlo is
competitive with parallel-tempering Monte Carlo for doing large-scale
spin-glass simulations at low but nonzero temperatures where
thermalization is difficult. Not only is population-annealing Monte
Carlo competitive in comparison to parallel-tempering Monte Carlo when
it comes to speed and statistical errors, it has the added benefits that
the free energy is readily accessible, multiple boundary conditions can
be simulated at the same time, the position of the temperatures in the
anneal schedule does not need to be tuned with as much care as in
parallel tempering, and it is trivially parallelizable on multi-core
architectures.

It is well known that parallel tempering is more efficient at finding
spin-glass ground states than simulated annealing
\cite{RoRiRaNiVo09,formula} because parallel tempering is more efficient
at overcoming free-energy barriers.  Here we find that population
annealing is comparably efficient to parallel tempering Monte Carlo and,
thus, also more efficient than simulated annealing. Nonetheless, because
of the strong similarities between population annealing and simulated
annealing, a detailed comparison of the two algorithms is informative
and sheds light on the importance of staying near equilibrium, even for
heuristics designed to find ground states.

The outline of the paper is as follows. We first introduce our benchmark
problem, the Edwards-Anderson Ising spin glass, and the population
annealing algorithm in Sec.~\ref{EAPAMC}.  We then study the properties of
population annealing for finding ground states of the Edwards-Anderson
model and compare population annealing with simulated annealing in
Sec.~\ref{CWS}.  We conclude by comparing the efficiency of
population annealing and parallel tempering in Sec.~\ref{CWP} and
present our conclusions in Sec.~\ref{conclusion}.

\section{Models and Methods}
\label{EAPAMC}
\subsection{The Edwards-Anderson model}

The Edwards-Anderson (EA) Ising spin-glass Hamiltonian is defined by
\begin{equation}
\mathcal{H}=-\sum\limits_{\langle ij \rangle} J_{ij} s_i s_j,
\end{equation}
where $s_i=\pm1$ are Ising spins on a $d$-dimensional hypercubic lattice
with periodic boundary conditions of size $N = L^3$. The summation
$\langle ij \rangle$ is over all nearest neighbor pairs. The couplings
$J_{ij}$ are independent Gaussian random variates with mean zero and
variance one. For Gaussian disorder, with probability one, there is a
unique pair of ground states for any finite system. We call a
realization of the couplings $\{J_{ij}\}$ a sample. Here we study the
three-dimensional (3D) EA model.

\subsection{Population Annealing and Simulated Annealing}
\label{sec:alg}

Population annealing (\pa) and simulated annealing (\sa) are closely
related algorithms that may be used as heuristics to find ground states
of the EA model. Both algorithms change the temperature of a system
through an annealing schedule from a high temperature where the system
is easily thermalized to a sufficiently low temperature where there is a
significant probability of finding the system in its ground state. At
each temperature in the annealing schedule $N_S$ sweeps of a
Markov-chain Monte Carlo algorithm are applied at the current
temperature. The annealing schedule consists of $N_T$ temperatures.  The
ground state is identified as the lowest-energy configuration
encountered during the simulation.

Population annealing differs from simulated annealing in that a
population of $R$ replicas of the system are cooled in parallel. At each
temperature step there is a resampling step, described below.

The resampling step in \pa~keeps the population close to the Gibbs
distribution as the temperature is lowered by differentially reproducing
replicas of the system depending on their energy.  Lower-energy replicas
may be copied several times and higher-energy replicas eliminated from
the population.  Consider a temperature step in which the temperature
$T = 1/\beta$
is lowered from from $\beta$ to $\beta^\prime$, where
$\beta^\prime>\beta$.  The reweighting factor required to transform the
Gibbs distribution from $\beta$ to $\beta^\prime$ for replica $i$ with
energy $E_i$ is $e^{-(\beta^\prime-\beta)E_i}$.  The expected number of
copies $\tau_i(\beta,\beta^\prime)$ of replica $i$ is proportional to
the reweighting factor
\begin{equation}
\tau_i(\beta,\beta^\prime)
 = \dfrac{e^{-(\beta^\prime-\beta)E_i}}{Q(\beta,\beta^\prime)},
\end{equation}
where $Q$ is a normalization factor that keeps the expected population 
size fixed,
\begin{equation}
\label{eq:Q}
Q(\beta,\beta^\prime)
 = \frac{1}{R}\sum\limits_{i=1}^{R_{\beta}} e^{-(\beta^\prime-\beta)E_i}.
\end{equation}
Here $R_{\beta}$ is the actual population size at inverse temperature
$\beta$. A useful feature of \pa~is that the absolute free energy can
be easily and accurately computed from the sum of the logarithm of the
normalizations $Q$ at each temperature step. Details can be found in
Ref.~\cite{A}.

Resampling is carried out by choosing a number of copies to make for
each replica in the population at $\beta^\prime$. There are various ways
to choose the integer number of copies $n_i(\beta,\beta^\prime)$ having
the correct (real) expectation $\tau_i(\beta,\beta^\prime)$.  The
population size can be fixed ($R_\beta=R$)  by using the multinomial
resampling \cite{B} or  residual resampling \cite{math}. Here we allow
the population size to fluctuate slightly and use {\em nearest-integer}
resampling. We let the number of copies be
$n_i(\beta,\beta^\prime)=\lfloor \tau_i(\beta,\beta^\prime) \rfloor$
with probability $\lceil\tau_i(\beta,\beta^\prime) \rceil -
\tau_i(\beta,\beta^\prime)$ and $n_i(\beta,\beta^\prime)=\lceil
\tau_i(\beta,\beta^\prime) \rceil$, otherwise. $\lfloor x \rfloor$ is
the greatest integer less than $x$ and $\lceil x \rceil$ is least
integer greater than $x$.  This choice insures that the mean of
$n_i(\beta,\beta^\prime)$ is $\tau_i(\beta,\beta^\prime)$ and the
variance is minimized.

The resampling step ensures that the new population is representative of
the Gibbs distribution at $\beta^\prime$ although for finite population
$R$, biases are introduced because the low-energy states are not fully
sampled. In addition, the population is now correlated due to the
creation of multiple copies.  Both of these problems are partially
corrected by carrying out Metropolis sweeps and the undersampling of
low-energy states is reduced by increasing $R$.  Indeed, for \pa, which
is an example of a sequential Monte Carlo method \cite{DoFrGo01},
systematic errors are eliminated in the large-$R$ limit.  By contrast,
for \pt, which is a Markov-chain Monte Carlo method, such systematic
errors are eliminated in the limit of a large number of Monte Carlo
sweeps.

In all our PA and SA simulations, the annealing schedule consists of
temperatures that are evenly spaced in $\beta = 1/T$ with the highest
temperature $1/T = \beta=0$ and the lowest temperature $1/T = \beta=5$.
The Markov chain Monte Carlo is the Metropolis algorithm and, unless
otherwise stated, there are $N_S=10$ Metropolis sweeps at each
temperature.

For both \pa~and \sa~the ground state is presumed to be the lowest
energy spin configuration encountered at the end of the simulation.  For
\sa~it is most efficient to do {\em multiple} runs and
choose the lowest energy from among the runs rather than do one very
long run. Thus, the \sa~results are typically stated as a function of
the number of runs $R$. Population annealing is inherently parallel and
we report results for a {\em single} run with a population size $R$.
Indeed, choosing the minimum energy configuration among $R$ runs of
\sa~is equivalent to running \pa~with the same population size but with
the resampling step turned off, which justifies using the same symbol
$R$ to describe the population size in \pa~and the number of runs in \sa.

While population annealing is primarily designed to sample from the
Gibbs distribution at nonzero temperature, here we are interested in its
performance for finding ground states. We test the hypothesis that the
resampling step in \pa~improves ground-state searches as compared to
\sa. The  motivation for this hypothesis is that the resampling step
removes high-energy spin configurations and replaces them with
low-energy configurations, thus potentially increasing the probability
of finding the ground state for a given value of $R$.

The equilibration of population annealing can be quantified using the
{\em family entropy} of the simulation. A small fraction of the initial
population has descendants in the final population at the lowest
temperature. Let $\nu_i$ be the fraction of the final population at the
lowest temperature descended from replica $i$ in the initial population.
Then the family entropy $S_f$ is given by
\begin{equation}
\label{eq:family}
S_f = -\sum_i \nu_i \log \nu_i .
\end{equation}
The exponential of the family entropy is an effective number of
surviving families. A high family entropy indicates smaller statistical
and systematic errors and can be used as a thermalization criterion
for the method.

\subsection{Measured quantities}

To compare \pa~and \sa~we investigated the following quantities.   For
\pa~let $\fGS(R)$ be the fraction of the population in the ground state
for a run with population size $R$. It is understood that $\fGS$ is
measured at the lowest simulated temperature. Clearly, the quantity
$\fGS(1)$ is simply the probability of finding the ground state in a
single run of \sa. Let $\pGS(R)$ be the probability of finding the
ground state in a run with population size $R$.  For \sa, $\pGS_{\rm
SA}(R)$ is the probability of finding the ground state in $R$
independent runs, i.e., 
\begin{equation}
\label{eq:psa}
\pGS_{\rm SA}(R)=1-[1-\fGS(1)]^R .
\end{equation}
However, for \pa~the resampling step tends to reproduce discoveries of
the ground state so that the probability $\pGS_{\rm PA}(R)$ is less than
the result for $R$ independent searches. What we actually measured is
$N_{\GS}$, the number of occurrences of the ground state in the
population from which we obtained $\fGS(R)=N_{\GS}/R$ in the case of
\pa~and $\fGS(1)$ in the case of \sa.

In the limit of large $R$, \pa~generates an equilibrium population
described by the Gibbs distribution so
\begin{equation}
\lim_{R \rightarrow \infty} \fGS(R) = \fGS_{\GS} ,
\end{equation} 
where $\fGS_{\GS}$ is the fraction of the ensemble
in the ground state,
\begin{equation}
\label{eq:fgs}
\fGS_{\GS} 	= \frac{1}{Z(\beta)}2e^{-\beta \eGS} 
		= 2 e^{-\beta \eGS+ \beta F(\beta)},
\end{equation}
where $\eGS$ is the ground-state energy, $Z(\beta)$ is the partition
function and $F(\beta) = -\log[Z(\beta)]/\beta$ is the Helmholtz free
energy.  As explained in Refs.~\cite{A,WaMaKa14,pamc}, \pa~yields
accurate estimates for the Helmholtz free energy based on the
normalization factors $Q(\beta,\beta^\prime)$ defined Eq.~\eqref{eq:Q}.
Thus, we have an independent prediction for the limiting value of
$\fGS(R)$.

We considered two disorder-averaged quantities as well.  The first is the
probability of finding the ground state, averaged over disorder samples,
\begin{equation}
\label{eq:eta}
\etaGS=\overline{\pGS},
\end{equation}
where the overbar indicates a disorder average. The quantity $\etaGS$ is
the primary measure we used to compare the three algorithms.

The second quantity, $\alpha$, is a disorder-averaged measure of accuracy
of finding the ground-state energy, i.e.,
\begin{equation}
\label{eq:alpha}
\alpha=1 - \overline{(E_{\rm min}/\eGS)} ,
\end{equation}
where $E_{\rm min}$ is the minimum energy found in the simulation, which
might {\em not} be the true ground-state energy $\eGS$.

\section{Results}

\subsection{Finding ground states with population annealing}

To compare population annealing and simulated annealing, we need a
collection of samples with known ground-state energies. In
Ref.~\cite{WaMaKa14} we reported on a simulation of approximately $5000$
samples of the 3D EA spin glass for size $L=4$, $6$, $8$, and $10$ using
large population runs of \pa.  Note that these sizes are typical of
recent ground-state studies of spin glasses.  Ground-state energies were
obtained from these runs by taking the lowest energy encountered in the
population at the lowest temperature, $\beta =5$, using
more-than-adequate resources. We used the data from this large-scale
simulation as the reference ground-state energy for each sample and
compared the same set of samples for smaller \pa~runs and for \sa~runs.
The population size and number of temperature steps in the reference
data set are shown in Table \ref{fgs}.  Our \pa implementation uses
\texttt{OPENMP} and each simulation runs on eight cores.

\begin{table}[h]
\caption{
Simulation parameters of the reference simulations of
Ref.~\cite{WaMaKa14} from which ground states were obtained. $L$ is the
linear system size, $R$ is the population size, $N_T$ is the number of
temperatures in the annealing schedule, $\min(N_{\GS})$ is the minimum
with respect to samples of the number of replicas in the ground state. }
\label{fgs}
\begin{tabular*}{\columnwidth}{@{\extracolsep{\fill}} l c c c c r }
	\hline
	\hline
	$L$ &$R$ &$N_T$ &$\min(N_{\GS})$ \\ \hline
	4   &5$\times10^4$ &101 &3370 \\
	6   &2$\times10^5$ &101 &1333 \\
	8   &5$\times10^5$ &201 &172 \\
	10  &1$\times10^6$ &301 &2 \\ \hline
	\hline
\end{tabular*}
\end{table}

Population annealing, like simulated annealing and parallel tempering,
is a heuristic method and it is not guaranteed to find the ground state
except in the limit of an infinite population size. Nonetheless, we have
confidence that we have found the ground state for all or nearly all
samples. For an algorithm like \pa~that is designed to sample the Gibbs
distribution at low temperature, the question of whether the true ground
state has been found is closely related to the question of whether
equilibration has been achieved at the lowest simulated temperature.  The
candidate ground state is defined as the minimum energy state in the
population at the lowest temperature $\beta$.  For an equilibrium
ensemble, the fraction of the ensemble in the ground $\fGS_{\GS}$ is
given by the Gibbs distribution, Eq.~\eqref{eq:fgs}.  If the number of
copies of the found ground state in the low-temperature population
$N_{\GS}$ is large and if the population is in equilibrium, then it is
unlikely that the true ground-state energy has not been found. Because,
if we have not found the true ground state, the  number of
copies of the true ground state, $R \fGS_{\GS}$, would be expected to be
even larger than $N_{\GS}$. Thus, if we believe the population is in
equilibrium at low temperature and if the candidate ground state is
found many times in the low-temperature population, then we have high
confidence that the candidate is the true ground state.

Of course, it  cannot be guaranteed that the population generated by
\pa~is in equilibrium at low temperature. However, the production runs
from which we  measured  ground-state energies  passed a
stringent thermalization test. We required a large effective number of
independent families  based on the family entropy, defined in
Eq.~\eqref{eq:family}.  We required $e^{S_f} \geq 100$; additional runs
were done for those samples that did not meet these criteria.  We also
compared our results for the same set of samples to results reported in
Ref.~\cite{burcu} using parallel tempering Monte Carlo.  We found good
agreement for both averaged quantities and the overlap distribution for
individual samples.

\begin{figure}
\begin{center}
\includegraphics[scale=0.9]{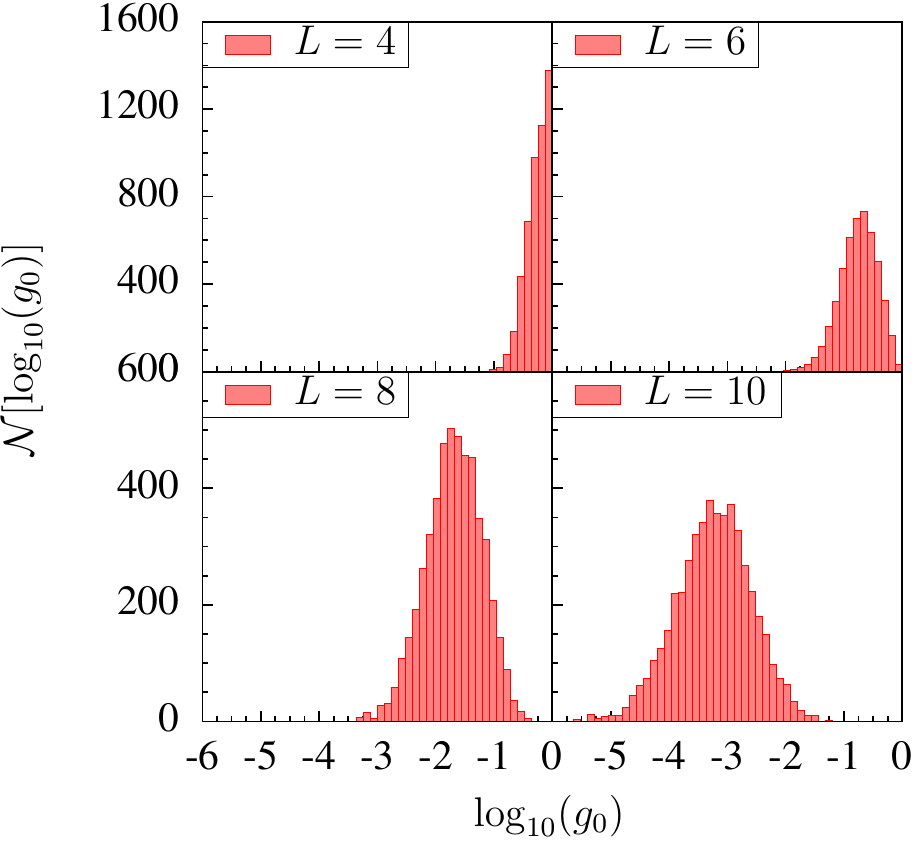}
\caption{(Color online)
Histogram of the number of samples with fraction in the ground state
$\fGS_{\GS}$ at $\beta=5$ for various sizes $L$, estimated from the
reference runs described in Table \ref{Nf}. $\mathcal
N[\log_{10}(\fGS_{\GS})]$ is the number of samples in the logarithmic
bin centered at $\log_{10}(\fGS_{\GS})$. There are a total of $50$ bins.
Note that as $L$ increases, the histograms shift rapidly to smaller
values. }
\label{Nf}
\end{center}
\end{figure}

In addition to the equilibration test, we recorded the number of copies
of the ground state in the population at the lowest temperature and
found that for most samples this number is large.  A histogram of
$N_{\GS}/R=\fGS(R) \approx \fGS_{\GS}$ of all samples is given in
Fig.~\ref{Nf} for each system size $L$. The minimum value of $N_{\GS}$
for each system size is shown in Table~\ref{fgs}. For the small fraction
($0.7$\%) of $L=10$ samples with $N_{\GS} <10$ we re-ran \pa~with a
$10$-fold larger population, $R =10^7$.  In no case did the ground-state
energy change.  In addition, for the one sample with $N_{\GS}=2$ we
confirmed the ground state using an exact branch and cut algorithm run
on the University of Cologne Spin Glass Server \cite{sgserver}.  Based
on the strict equilibration criteria and the large number of ground states
reported in Table~\ref{fgs}, we are confident that we have found true
ground states for all samples.

As an additional check, we compared the disorder-averaged ground-state
energy per spin against values in the literature using the hybrid
genetic algorithm \cite{gse} and parallel tempering (PT)
\cite{RoRiRaNiVo09}.  The comparison is shown in Table~\ref{gse} and
reveals that all three methods yield the same average energy within
statistical errors.

\begin{table}[h]
\caption{
Comparison of the disorder averaged ground-state energy per spin for the
EA model with those obtained from the hybrid genetic
algorithm~\cite{gse} and \pt~\cite{RoRiRaNiVo09}.
}
\label{gse}
\begin{tabular*}{\columnwidth}{@{\extracolsep{\fill}} l c c c }
	\hline
	\hline
	$L$ &\pa &Hybrid genetic &\pt \\ \hline
	4 &-1.6639(14) &-1.6655(6) &-1.6660(2) \\
	6 &-1.6899(7) &-1.6894(5) &-1.6891(4) \\ 
	8 &-1.6961(5) &-1.6955(4) &-1.6955(6) \\ 
	10 &-1.6980(3) &-1.6975(5) &-1.6981(7) \\ \hline
	\hline
\end{tabular*}
\end{table}

A striking feature of Fig.~\ref{Nf} is that the fraction of the ensemble
in the ground state $\fGS_{\GS}$ decreases rapidly as $L$ increases.
Thus, for any temperature-based heuristic, including \pa, \sa, and \pt,
it is necessary to simulate at lower temperatures and/or use larger
populations (or, for \pt, longer runs) as $L$ increases. To understand
this requirement more formally we rewrite Eq.~\eqref{eq:fgs} in terms of
intensive quantities
\begin{equation}
\fGS_{\GS}=2\exp[-N\beta (e_{\GS}-f(\beta))],
\end{equation}
where $e_{\GS}$ and $f(\beta)$ are the ground-state energy and free
energy per spin, respectively, and $N=L^3$ is the number of spins. In
the thermodynamic limit, $[e_{\GS}-f(\beta)]$ is expected to converge to
a positive number that is independent of the disorder realization.
Thus, for fixed $\beta$, the fraction of the ensemble in the ground
state decreases exponentially in the system size.

As discussed in Sec.~\ref{sec:alg}, population annealing gives a direct
estimator of the free energy, thus we can independently measure all of
the quantities in Eq.~\eqref{eq:fgs} and carry out a disorder average.
Because the observables on the right-hand side of Eq.~\eqref{eq:fgs}
appear in the exponent, it is convenient to take the logarithm and then
carry out the disorder average. Table \ref{gsmodel} compares
$\overline{\log_{10} \fGS_{\GS}}$ and
$\log_{10}2-\beta\overline{(E_{\GS}-F)}/\log(10)$ at $\beta=5$.  The
table confirms the expected equilibrium behavior of the fraction in the
ground state. Note that the observables $\fGS_{\GS}$, $E_{\GS}$, and $F$
are not entirely independent quantities, which explains why the
statistical errors are significantly larger than the difference in the
values.  On the other hand, if the simulation was not in thermal
equilibrium, these quantities would not agree.

\begin{table}[h]
\caption{
Comparison of the disorder average of the log of the two sides of 
Eq.~\eqref{eq:fgs} at $\beta=5$.
}
\label{gsmodel}
\begin{tabular*}{\columnwidth}{@{\extracolsep{\fill}} l c c }
	\hline
	\hline
	$L$ &$\overline{\log_{10}\fGS_{\GS}}$ &$\log_{10}2-\beta\overline{(E_{\GS}-F)}/\log(10)$ \\ \hline
	4 &-0.2644(28) &-0.2643(28) \\
	6 &-0.7573(46) &-0.7563(46) \\ 
	8 &-1.6933(77) &-1.6925(67) \\ 
	10 &-3.2358(104) &-3.2297(91) \\ \hline
	\hline
\end{tabular*}
\end{table}

For all the reasons discussed above we believe that we have found the
true ground state for all samples.  However, our main conclusions would
not be affected if a small fraction of the reference ground states are
not true ground states.

\subsection{Comparison between population annealing and simulated annealing}
\label{CWS}

\subsubsection{Detailed comparison for a single sample}
\label{single}

In this section we present a comparison of population annealing and
simulated annealing for a single disorder realization.  This sample was chosen to be hardest to equilibrate for $L=8$ based on having the
smallest family entropy [see Eq.~\eqref{eq:family}]; however, it has a
probability of being in the ground state at the lowest temperature near
the average for size $L=8$. For this sample we  confirmed the
ground-state energy found in the reference \pa~run using the University
of Cologne Spin Glass Server \cite{sgserver}.

\begin{figure}[htb]
\begin{center}
\includegraphics[scale=0.9]{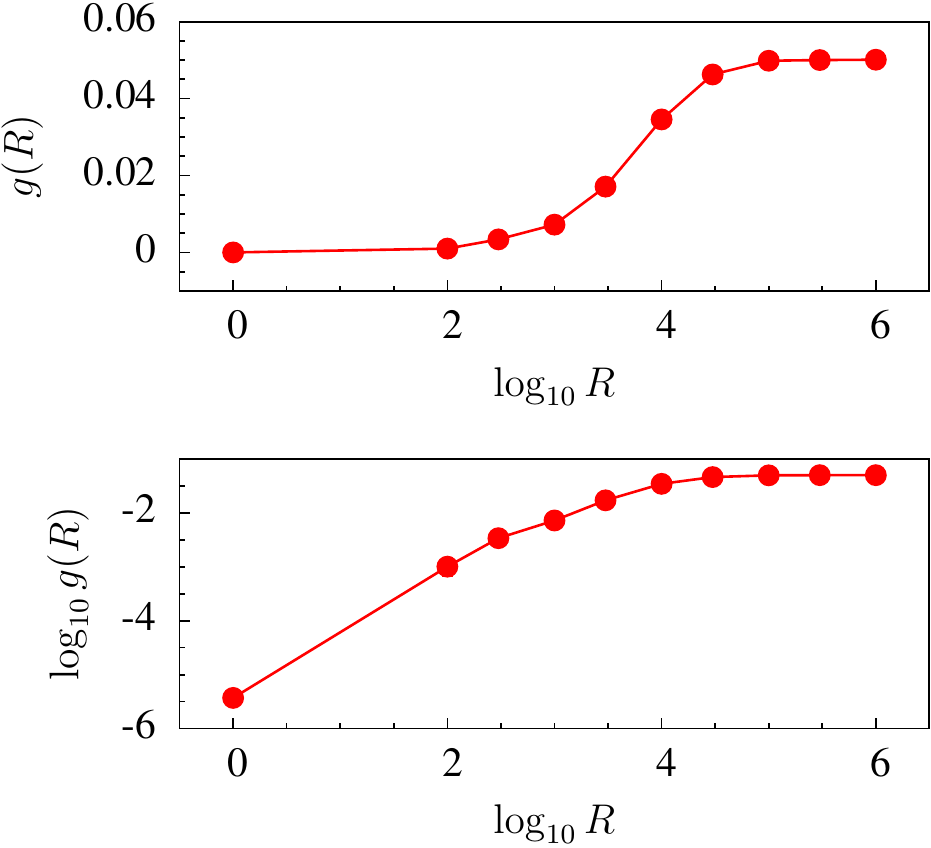}
\caption{(Color online) 
The fraction of the population in the ground state $\fGS(R)$ as a
function of population size $R$  for a single sample using \pa~with
$N_T=101$ and $N_S=10$. The point at $\log_{10}R=0$ corresponds to the
probability that a single run of \sa~will yield the ground state. The
upper panel is a log-linear plot and the lower panel is a log-log plot
of the same data. Error bars are smaller than the symbols.}
\label{fr}
\end{center}
\end{figure}

Figure~\ref{fr} shows the fraction of the population in the ground state
$\fGS(R)$ as a function of population size $R$ for \pa. The result for
the probability that \sa~has found the ground state in a single run is
simply the value at $R=1$. In this simulation, we used $N_T=101$
temperatures with $N_S=10$ sweeps per temperature for both algorithms.
It is striking that the fraction of ground states in the population
increases by about four orders of magnitude from the small value for
\sa, $\fGS(1)$ to the limiting value  for \pa~for large $R$, $\fGS(10^6)
\approx \fGS_{\GS}$.  This result shows that resampling greatly
increases the probability that a member of the \pa~population is in the
ground state. It suggests that even though equilibration is not required
for finding ground states, the probability of finding the ground state
is improved when the simulation is maintained near thermal equilibrium.
Of course, remaining near equilibrium as the temperature is lowered is
also a motivation for \sa~but lacking the resampling step, \sa~falls out
of equilibrium once the free-energy landscape roughness significantly
exceeds $k_{\rm B}T$. However, the ratio of $\fGS(R)/\fGS(1)$ is an
overestimate of the ratio the probabilities for actually finding the
ground state for a fixed $R$ because once the ground state is discovered
in \pa, it is likely to be reproduced many times.

\begin{figure}
\begin{center}
\includegraphics[scale=0.9]{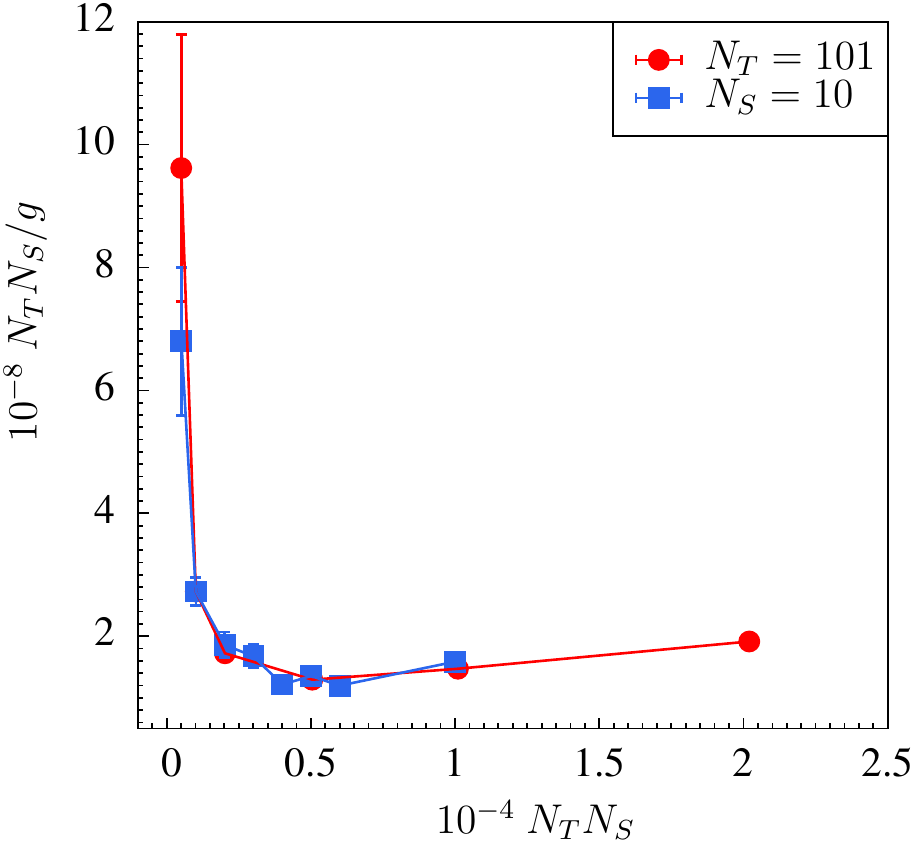}
\caption{(Color online) 
The computational work divided by the probability of finding the ground
state in a single \sa~run, $N_TN_S/\fGS$ vs the computational work
$N_TN_S$ for a single sample.  The two curves correspond to holding
$N_S=10$ fixed and varying $N_T$ (blue squares) and holding $N_T=101$
fixed and varying $N_S$ (red circles).  Smaller values of $N_TN_S/\fGS$
correspond to more efficient simulations.}
\label{NTNS}
\end{center}
\end{figure}

The probability of finding the  ground state $\mathcal{P}$ for a given
amount of computational work is an appropriate metric to compare the two
algorithms. We measured the amount of work $W$ in Metropolis sweeps,
$W=RN_TN_S$. In most of our comparisons we used the same value of $N_T$
and $N_S$ for both \pa~and \sa.  However, it is not clear whether the
two algorithms are optimized with the same values of $N_T$ and $N_S$.
We performed additional optimization of \sa~varying $N_T$ and $N_S$.  We
used the computational work divided by the probability of finding the
ground state in a single \sa~run, with $N_TN_S/\fGS$ as a figure of merit.
Note that in the relevant large-$R$ regime, minimizing $N_TN_S/\fGS$ is
equivalent to maximizing $\mathcal{P}$ for a fixed amount of work.
Figure \ref{NTNS} shows $N_TN_S/\fGS$ versus $N_TN_S$ and reveals a broad
minimum near $N_TN_S \approx 5 \times 10^3$.  We therefore performed
\sa~simulations at the same value used for \pa, $N_TN_S=1010$, and a more
nearly optimal value, $N_TN_S=5000$.  Note that for \sa~it is only the
product, $N_TN_S$, that determines the efficiency, not $N_T$ and $N_S$
separately.  Note also that the efficiency decreases when  $N_TN_S$ is
too large, suggesting that it is better to do many shorter \sa~runs
rather than a single long run.

\begin{figure}[htb]
\begin{center}
\includegraphics[scale=0.9]{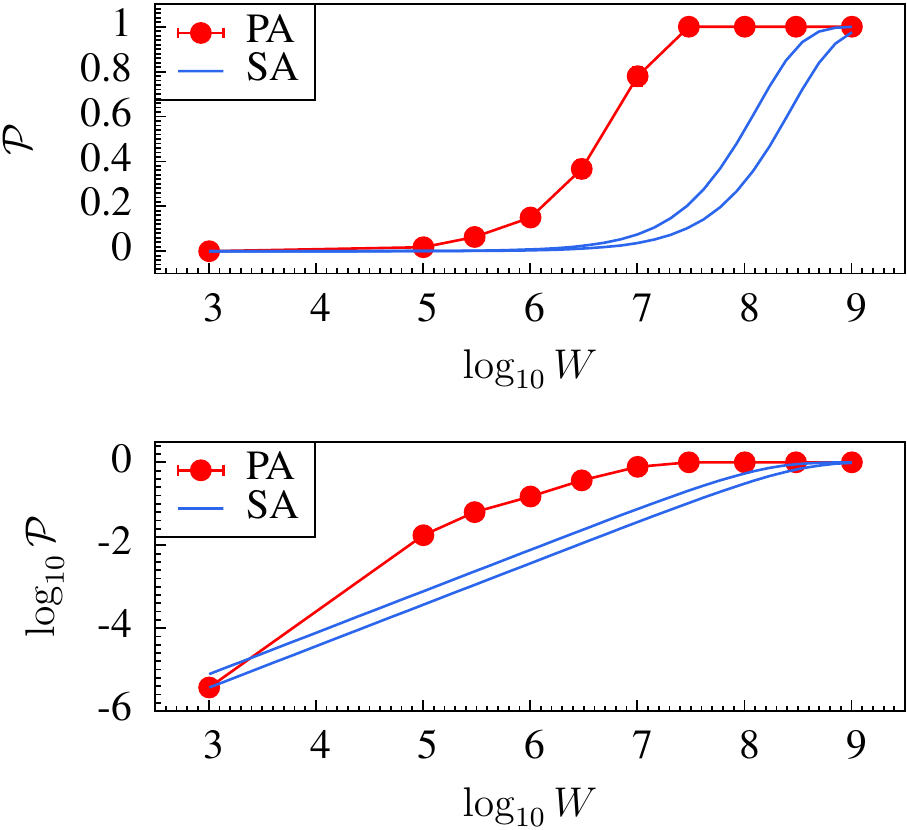}
\caption{(Color online) 
Probability of finding the ground state $\mathcal{P}$ as a function
of the computational work $W=R N_TN_S$ for a single sample for both
\sa~and \pa. The computational work is varied by changing population size
$R$, holding $N_TN_S$ fixed. For \pa~and the lower \sa~curve,
$N_TN_S=1010$ while for the upper \sa~curve,  $N_TN_S=5000$,  which is
near the optimum value for \sa. The upper panel is a log-linear plot
and the lower panel is a log-log plot. Error bars for \pa~are smaller
than the symbols. The \sa~curves are obtained from Eq.~\eqref{eq:psa}.}
\label{frreal}
\end{center}
\end{figure}

Figure~\ref{frreal} compares $\mathcal{P}_{\rm SA}$, obtained from
Eq.~\eqref{eq:psa}, and $\mathcal{P}_{\rm PA}(R)$, obtained from multiple
runs of \pa~as a function of the computational work $W$.  In this
simulation we used $N_T=101$ temperatures with $N_S=10$ for \pa~and the
lower \sa~curve. The upper  \sa~curve  corresponds to the optimal value
$N_TN_S=5000$.    Computational work was varied by changing $R$ holding
$N_T$ and $N_S$ fixed.  For intermediate values of $R$, corresponding to
realistic simulations, $\mathcal{P}_{\rm PA}$ exceeds $\mathcal{P}_{\rm
SA}$ by one or two orders of magnitude and the amount of work needed to
be nearly certain of finding the ground is also more than an order of
magnitude less for \pa~than \sa.  Note that the effect of optimizing
\sa~is only about a factor of $2$.  We conclude that for this sample,
there is a large difference in efficiency between \pa~and \sa~and this
difference cannot be explained by a difference in the optimization of
the two methods.  To see whether this difference is typical and how it
depends on system size, in Sec.~\ref{large} we consider averages over
disorder realizations.

\subsubsection{Disorder-averaged comparison}
\label{large}

We  compared population annealing and simulated annealing for
approximately $5000$ disorder realizations for each of the four system
sizes, $L=4$, $6$, $8$, and $10$, and for several different population
sizes.  For \sa~the population size refers to the number of independent
runs. Both algorithms use the same annealing schedule with evenly
spaced inverse temperatures starting with infinite temperature and
ending at $T_0=0.2$. The number of sweeps per temperature is $N_S=10$.
The population sizes $R$, number of temperatures in the annealing
schedule $N_T$, the number of disorder realizations $M$ and the
corresponding parameters for the reference runs are given in
Table~\ref{parameters}.

\begin{table}[h]
\caption{
Parameters of the numerical simulations for comparison between \pa~and
\sa. $R$ is the population size, $N_T$ is the number of temperatures,
and $M$ is the number of samples studied. The reference parameters are
for the \pa~runs used to estimate the ground-state energy for each
sample.}
\label{parameters}
\begin{tabular*}{\columnwidth}{@{\extracolsep{\fill}} l c c c c r }
	\hline
	\hline
	$L$ &$\log_{10}R$ &$N_T$ &$M$ &Ref. $R$ &Ref. $N_T$ \\
	\hline
	4  & \{1,2,3,4\}   &101 &4941 &5$\times10^4$ &101 \\
	6  & \{1,2,3,4,5\} &101 &4959 &2$\times10^5$ &101 \\
	8  & \{1,2,3,4,5\} &101 &5099 &5$\times10^5$ &201 \\
	10 & \{1,2,3,4,5\} &201 &4945 &1$\times10^6$ &301 \\ \hline
	\hline
\end{tabular*}
\end{table}

\begin{figure}[htb]
\begin{center}
\includegraphics[scale=0.9]{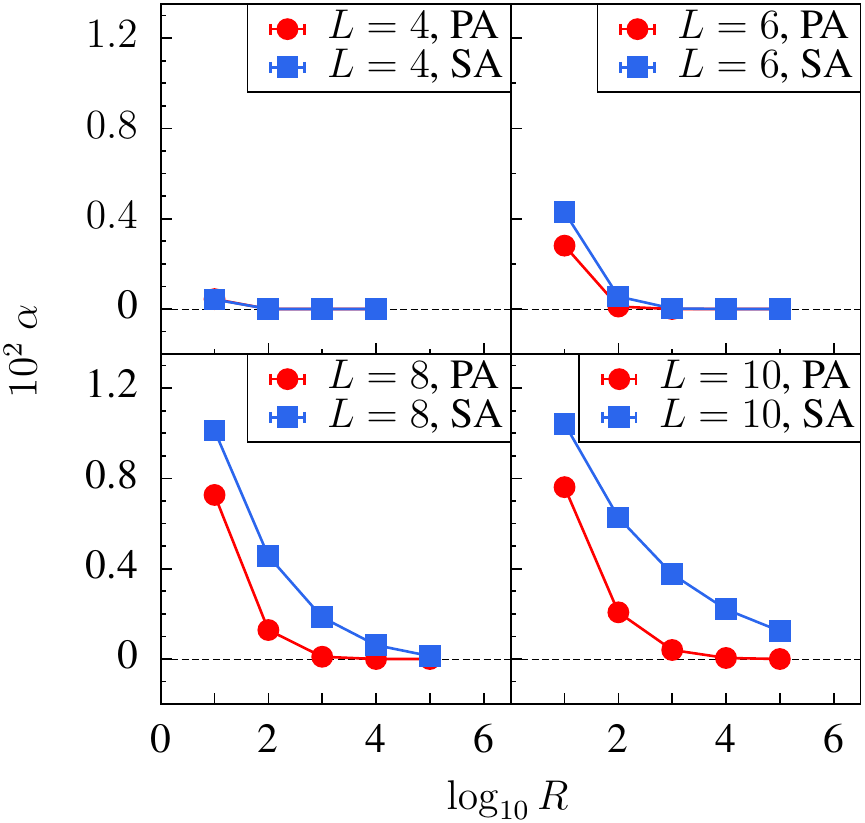}
\caption{(Color online) 
Error in approximating the the ground-state energy ($\alpha$) vs log
population size, $\log_{10}(R)$.}
\label{alpha}
\end{center}
\end{figure}

Figure~\ref{alpha} shows $\alpha$, the disorder averaged error in
finding the ground state [see Eq.~\eqref{eq:alpha}], as function of
population size $R$ for \sa~and \pa. For small systems neither algorithm
makes significant errors even for small populations but as the system
size increases, \pa~is significantly more accurate.

\begin{figure}[htb]
\begin{center}
\includegraphics[scale=0.9]{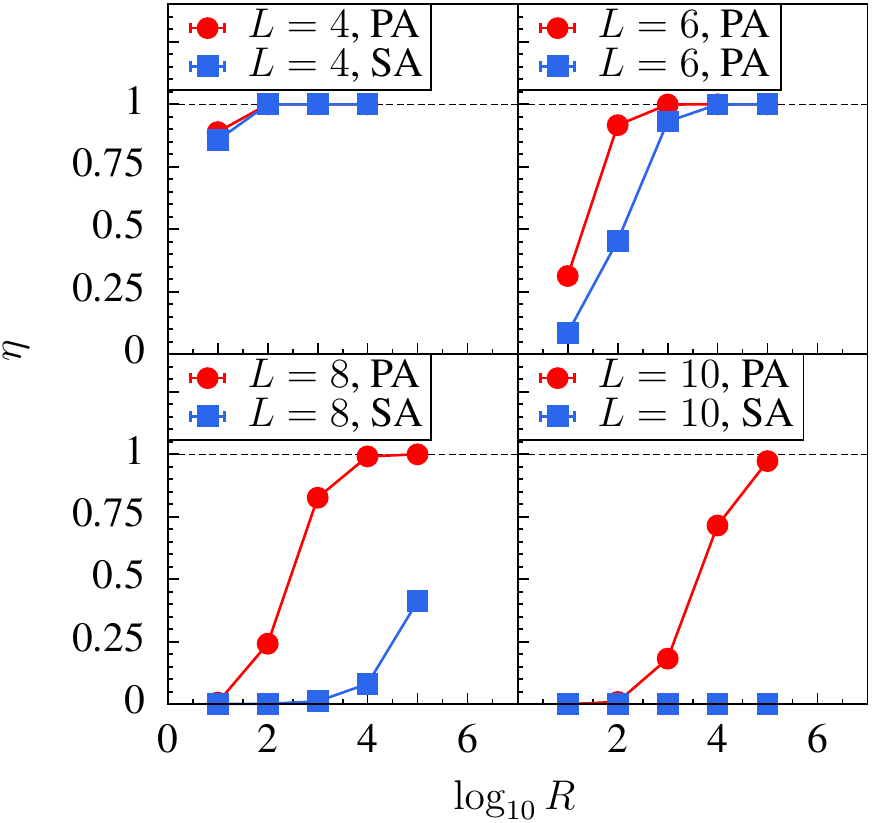}
\caption{(Color online) 
Fraction of samples for which the ground state is found ($\etaGS$) vs log
population size, $\log_{10}(R)$ for population annealing and simulated
annealing.}
\label{beta}
\end{center}
\end{figure}

Figure~\ref{beta} shows $\etaGS$, the disorder-averaged fraction of
samples for which the ground state is found [see Eq.~\eqref{eq:eta}], as
a function of population size $R$.  Again, we see that for $L= 4$ and
$6$, the two algorithms are comparable but for $L=8$ and $10$,
population annealing is far more likely to find the ground state for the
same population size.  It is clear from Figs.~\ref{alpha} and \ref{beta}
that population annealing is both more accurate and more efficient at
finding ground states than simulated annealing and that as system size
increases, the relative advantage of \pa~over \sa~increases.

\section{Comparison between Population Annealing and Parallel Tempering}
\label{CWP}

In this section, we compare the efficiency of population annealing (\pa)
and parallel tempering (\pt) when finding ground states.  We first
briefly describe parallel-tempering Monte Carlo.  Parallel tempering
simultaneously simulates $N_T$ replicas of the system at $N_T$ different
temperatures. In addition to single-temperature Metropolis sweeps,
\pa~uses {\em replica exchange} moves in which two replicas at
neighboring temperatures swap temperatures.  To satisfy detailed
balance, the swap probability $p_{\rm swap}$ is given by
\begin{equation}
\label{ }
p_{\rm swap} = 
\min\Big[1,\exp\left[(\beta^\prime - \beta)(E^\prime - E) \right]\Big] ,
\end{equation}
where $E$ and $E^\prime$ are the  energies of the replicas proposed for
exchange at temperatures $\beta$ and $\beta^\prime$, respectively.

Results for \pt~are taken from Rom\'a {\em et al.}~\cite{RoRiRaNiVo09},
who studied the disorder-averaged probability of finding the ground
state $\etaGS$ for the 3D EA model for the same sizes considered here.
They gave an empirical fit of their data of the form,
\begin{equation}
\label{eq:rfit}
\eta=\dfrac{e^{qx}}{1+e^{qx}},
\end{equation}
where $q$ is a fitting parameter and $x$ is a  function of the
computational work $W$ and system size $L$ defined as
\begin{equation}
\label{eq:rfit2}
x=[\log(W/2)-(bL^c-a)]/L^d,
\end{equation}
and the work is calculated in units of Monte Carlo sweeps. For \pt, the
computational work is given by $W=N_TN_S$ while for \pa it is given
by $W=RN_TN_S$. We assume that the work involved in replica exchange
moves for \pt~and in population resampling for \pa~is negligible
compared to the work associated with the Metropolis sweeps. The fitting
parameters for the 3D EA model reported in Ref.~\cite{RoRiRaNiVo09} are
$a=-0.05$, $b=1.55$, $c=1$, $d=0.2$, and $q=2$.

\begin{figure}[htb]
\begin{center}
\includegraphics[scale=0.9]{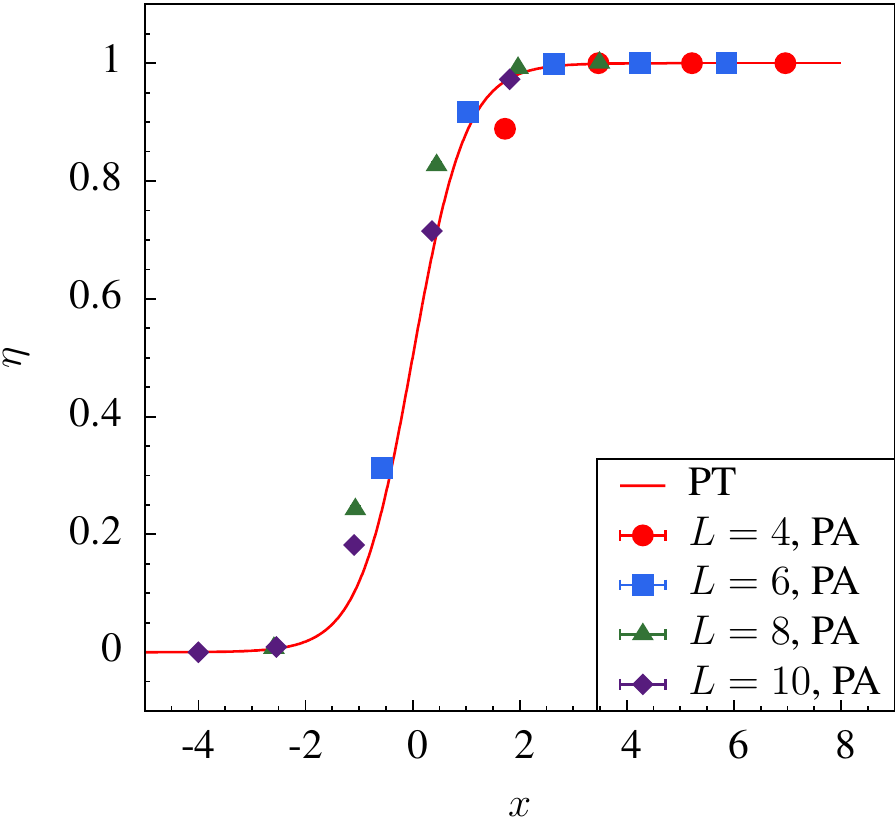}
\caption{(Color online) 
Fraction of samples for which the ground state is found ($\etaGS$) as a
function of the scaled computational work $x$ [Eq.~\eqref{eq:rfit2}] for
both population annealing and parallel tempering. The curve is taken
from the empirical fit [Eq.~\eqref{eq:rfit}] of Ref.~\cite{RoRiRaNiVo09}.}
\label{aa}
\end{center}
\end{figure}

Figure \ref{aa} shows $\etaGS$, the fraction of samples for which the
ground state is correctly found versus the scaled work $x$ for our \pa~data
(points) and the fit for \pt~from \cite{RoRiRaNiVo09} (solid curve).  It
is striking that both algorithms perform nearly identically over the
whole range of sizes and amounts of computational work.

\section{Conclusion}
\label{conclusion}

We have carried out a detailed comparison of three Monte Carlo
heuristics based on thermal annealing for finding ground states of
spin-glass Hamiltonians.  The algorithms compared are population-annealing, 
simulated-annealing and parallel-tempering Monte Carlo.  We
find that population annealing is more efficient than simulated
annealing and has better scaling with the system size.  In particular,
the CPU time needed for resampling the population is negligible. Thus,
with a similar numerical effort as for simulated annealing, population
annealing provides a sizable performance improvement.  

We  find that population annealing and parallel tempering are comparably
efficient for finding spin-glass ground states.  Population annealing,
however, is much better suited to a massively parallel implementation
and would be the preferred choice for large systems or when ground
states are required quickly.  A general conclusion is that Monte Carlo
heuristics based on thermal annealing are enhanced by mechanisms that
improve thermalization at every temperature.  In population annealing
this mechanism is resampling and in parallel tempering it is replica
exchange. Simulated annealing depends entirely on local Monte Carlo
moves and fails to remain close to equilibrium at low temperatures where
the free-energy landscape is rough. Furthermore, we observed that the
ensemble defined by simulated annealing has far less weight in the
ground state than the equilibrium ensemble for realistic computational
effort.  This deficiency results in a significantly lower probability of
finding the ground state for a given amount of computational effort as
compared to either population annealing or parallel tempering, which
stay close to thermal equilibrium.

There is no obvious reason to suppose that the temperature-dependent
Gibbs distribution is the best target distribution for improved
heuristics such as population annealing or parallel-tempering Monte
Carlo. Distributions other than the Gibbs distribution that concentrate
on the ground state as ``temperature'' is decreased might perform even
better than the Gibbs distribution and should be investigated.

\section*{Acknowledgments}

The work of J.M.~and W.W.~was supported in part from NSF Grant
No.~DMR-1208046. H.G.K.~acknowledges support from the NSF (Grant
No.~DMR-1151387).  H.G.K.~thanks Jeff Lebowski for introducing the
slacker formalism to tackle complex and frustrating problems.  We also
acknowledge HPC resources from Texas A\&M University (Eos cluster) and
thank ETH Zurich for CPU time on the Euler cluster.  Finally, we thank
F.~Liers and the University of Cologne Spin Glass Server for providing
exact ground-state instances.

\bibliographystyle{apsrevtitle}
\bibliography{references,refs}

\begin{thebibliography}{30}
\expandafter\ifx\csname natexlab\endcsname\relax\def\natexlab#1{#1}\fi
\expandafter\ifx\csname bibnamefont\endcsname\relax
  \def\bibnamefont#1{#1}\fi
\expandafter\ifx\csname bibfnamefont\endcsname\relax
  \def\bibfnamefont#1{#1}\fi
\expandafter\ifx\csname citenamefont\endcsname\relax
  \def\citenamefont#1{#1}\fi
\expandafter\ifx\csname url\endcsname\relax
  \def\url#1{\texttt{#1}}\fi
\expandafter\ifx\csname urlprefix\endcsname\relax\def\urlprefix{URL }\fi
\providecommand{\bibinfo}[2]{#2}
\providecommand{\eprint}[2][]{\url{#2}}

\bibitem[{\citenamefont{Stein and Newman}(2013)}]{book}
\bibinfo{author}{\bibfnamefont{D.~L.} \bibnamefont{Stein}} \bibnamefont{and}
  \bibinfo{author}{\bibfnamefont{C.~M.} \bibnamefont{Newman}},
  \emph{\bibinfo{title}{Spin Glasses And Complexity}}
  (\bibinfo{publisher}{Princeton University Press}, \bibinfo{year}{2013}).

\bibitem[{\citenamefont{Carter et~al.}(2002)\citenamefont{Carter, A.J.Bray, and
  Moore}}]{stiffness1}
\bibinfo{author}{\bibfnamefont{A.}~\bibnamefont{Carter}},
  \bibinfo{author}{\bibnamefont{A.J.Bray}}, \bibnamefont{and}
  \bibinfo{author}{\bibfnamefont{M.}~\bibnamefont{Moore}},
  \emph{\bibinfo{title}{{Aspect-Ratio Scaling and the Stiffness Exponent
  $\theta$ for Ising Sping Glasses}}}, \bibinfo{journal}{Phys. Rev. Lett.}
  \textbf{\bibinfo{volume}{88}}, \bibinfo{pages}{077201}
  (\bibinfo{year}{2002}).

\bibitem[{\citenamefont{Hartmann}(1999)}]{stiffness2}
\bibinfo{author}{\bibfnamefont{A.~K.} \bibnamefont{Hartmann}},
  \emph{\bibinfo{title}{{Scaling of stiffness energy for three-dimensional $\pm
  J$ Ising spin glasses}}}, \bibinfo{journal}{Phys. Rev. E}
  \textbf{\bibinfo{volume}{59}}, \bibinfo{pages}{84} (\bibinfo{year}{1999}).

\bibitem[{\citenamefont{Hukushima}(1999)}]{stiffness3}
\bibinfo{author}{\bibfnamefont{K.}~\bibnamefont{Hukushima}},
  \emph{\bibinfo{title}{{Domain-wall free energy of spin-glass models:
  Numerical method and boundary conditions}}}, \bibinfo{journal}{Phys. Rev. E}
  \textbf{\bibinfo{volume}{60}}, \bibinfo{pages}{3606} (\bibinfo{year}{1999}).

\bibitem[{\citenamefont{Barahona}(1982)}]{NPhard}
\bibinfo{author}{\bibfnamefont{F.}~\bibnamefont{Barahona}},
  \emph{\bibinfo{title}{{On the computational complexity of Ising spin glass
  models}}}, \bibinfo{journal}{J. Phys. A: Math. Gen.}
  \textbf{\bibinfo{volume}{15}}, \bibinfo{pages}{3241} (\bibinfo{year}{1982}).

\bibitem[{\citenamefont{Moore and Mertens}(2011)}]{MoMe11}
\bibinfo{author}{\bibfnamefont{C.}~\bibnamefont{Moore}} \bibnamefont{and}
  \bibinfo{author}{\bibfnamefont{S.}~\bibnamefont{Mertens}},
  \emph{\bibinfo{title}{The Nature of Computation}} (\bibinfo{publisher}{Oxford
  University Press}, \bibinfo{year}{2011}).

\bibitem[{\citenamefont{Trebst et~al.}(2006)\citenamefont{Trebst, Troyer, and
  Hansmann}}]{protein_folding}
\bibinfo{author}{\bibfnamefont{S.}~\bibnamefont{Trebst}},
  \bibinfo{author}{\bibfnamefont{M.}~\bibnamefont{Troyer}}, \bibnamefont{and}
  \bibinfo{author}{\bibfnamefont{U.~H.} \bibnamefont{Hansmann}},
  \emph{\bibinfo{title}{{Optimized parallel tempering simulations of
  proteins}}}, \bibinfo{journal}{J. Chem. Phys.}
  \textbf{\bibinfo{volume}{124}}, \bibinfo{pages}{174903}
  (\bibinfo{year}{2006}).

\bibitem[{\citenamefont{Simone et~al.}(1995)\citenamefont{Simone, Diehl,
  J\"unger, Mutzel, Reinelt, and Rinaldi}}]{Branch_and_cut}
\bibinfo{author}{\bibfnamefont{C.~D.} \bibnamefont{Simone}},
  \bibinfo{author}{\bibfnamefont{M.}~\bibnamefont{Diehl}},
  \bibinfo{author}{\bibfnamefont{M.}~\bibnamefont{J\"unger}},
  \bibinfo{author}{\bibfnamefont{P.}~\bibnamefont{Mutzel}},
  \bibinfo{author}{\bibfnamefont{G.}~\bibnamefont{Reinelt}}, \bibnamefont{and}
  \bibinfo{author}{\bibfnamefont{G.}~\bibnamefont{Rinaldi}},
  \emph{\bibinfo{title}{Exact ground states of {I}sing spin glasses: New
  experimental results with a branch and cut algorithm}}, \bibinfo{journal}{J.
  Stat. Phys.} \textbf{\bibinfo{volume}{80}}, \bibinfo{pages}{487}
  (\bibinfo{year}{1995}).

\bibitem[{\citenamefont{Hartmann and Rieger}(2001)}]{hartmann:01}
\bibinfo{author}{\bibfnamefont{A.~K.} \bibnamefont{Hartmann}} \bibnamefont{and}
  \bibinfo{author}{\bibfnamefont{H.}~\bibnamefont{Rieger}},
  \emph{\bibinfo{title}{Optimization Algorithms in Physics}}
  (\bibinfo{publisher}{Wiley-VCH}, \bibinfo{address}{Berlin},
  \bibinfo{year}{2001}).

\bibitem[{\citenamefont{Hartmann and Rieger}(2004)}]{hartmann:04}
\bibinfo{author}{\bibfnamefont{A.~K.} \bibnamefont{Hartmann}} \bibnamefont{and}
  \bibinfo{author}{\bibfnamefont{H.}~\bibnamefont{Rieger}},
  \emph{\bibinfo{title}{New Optimization Algorithms in Physics}}
  (\bibinfo{publisher}{Wiley-VCH}, \bibinfo{address}{Berlin},
  \bibinfo{year}{2004}).

\bibitem[{\citenamefont{B{\v a}utu and B{\v a}utu}(2008)}]{GA_and_BPSO}
\bibinfo{author}{\bibfnamefont{A.}~\bibnamefont{B{\v a}utu}} \bibnamefont{and}
  \bibinfo{author}{\bibfnamefont{E.}~\bibnamefont{B{\v a}utu}},
  \emph{\bibinfo{title}{Searching ground states of {I}sing spin glasses with
  genetic algorithms and binary particle swarm optimization}}, in
  \emph{\bibinfo{booktitle}{Nature Inspired Cooperative Strategies for
  Optimization (NICSO 2007)}}, edited by
  \bibinfo{editor}{\bibfnamefont{N.}~\bibnamefont{Krasnogor}},
  \bibinfo{editor}{\bibfnamefont{G.}~\bibnamefont{Nicosia}},
  \bibinfo{editor}{\bibfnamefont{M.}~\bibnamefont{Pavone}}, \bibnamefont{and}
  \bibinfo{editor}{\bibfnamefont{D.}~\bibnamefont{Pelta}}
  (\bibinfo{publisher}{Springer Berlin Heidelberg}, \bibinfo{year}{2008}), vol.
  \bibinfo{volume}{129} of \emph{\bibinfo{series}{Studies in Computational
  Intelligence}}, pp. \bibinfo{pages}{85--94}.

\bibitem[{\citenamefont{Boettcher and Percus}(2001)}]{BoPe01}
\bibinfo{author}{\bibfnamefont{S.}~\bibnamefont{Boettcher}} \bibnamefont{and}
  \bibinfo{author}{\bibfnamefont{A.~G.} \bibnamefont{Percus}},
  \emph{\bibinfo{title}{Optimization with extremal dynamics}},
  \bibinfo{journal}{Phys. Rev. Lett.} \textbf{\bibinfo{volume}{86}},
  \bibinfo{pages}{5211} (\bibinfo{year}{2001}).

\bibitem[{\citenamefont{Middleton}(2004)}]{Middleton04}
\bibinfo{author}{\bibfnamefont{A.~A.} \bibnamefont{Middleton}},
  \emph{\bibinfo{title}{Improved extremal optimization for the {I}sing spin
  glass}}, \bibinfo{journal}{Phys. Rev. E} \textbf{\bibinfo{volume}{69}},
  \bibinfo{pages}{055701} (\bibinfo{year}{2004}).

\bibitem[{\citenamefont{Kirkpatrick et~al.}(1983)\citenamefont{Kirkpatrick,
  Gelatt, and Vecchi}}]{SAMC}
\bibinfo{author}{\bibfnamefont{S.}~\bibnamefont{Kirkpatrick}},
  \bibinfo{author}{\bibfnamefont{C.~D.} \bibnamefont{Gelatt}},
  \bibnamefont{and} \bibinfo{author}{\bibfnamefont{M.~P.}
  \bibnamefont{Vecchi}}, \emph{\bibinfo{title}{Optimization by simulated
  annealing}}, \bibinfo{journal}{Science} \textbf{\bibinfo{volume}{220}},
  \bibinfo{pages}{671} (\bibinfo{year}{1983}).

\bibitem[{\citenamefont{Swendsen and Wang}(1986)}]{ptmc1}
\bibinfo{author}{\bibfnamefont{R.~H.} \bibnamefont{Swendsen}} \bibnamefont{and}
  \bibinfo{author}{\bibfnamefont{J.-S.} \bibnamefont{Wang}},
  \emph{\bibinfo{title}{Replica {M}onte {C}arlo simulations of spin glasses}},
  \bibinfo{journal}{Phys. Rev. Lett.} \textbf{\bibinfo{volume}{57}},
  \bibinfo{pages}{2607} (\bibinfo{year}{1986}).

\bibitem[{\citenamefont{Geyer}(1991)}]{ptmc2}
\bibinfo{author}{\bibfnamefont{C.}~\bibnamefont{Geyer}}, in
  \emph{\bibinfo{booktitle}{Computing Science and Statistics: 23rd Symposium on
  the Interface}}, edited by \bibinfo{editor}{\bibfnamefont{E.~M.}
  \bibnamefont{Keramidas}} (\bibinfo{publisher}{Interface Foundation},
  \bibinfo{address}{Fairfax Station}, \bibinfo{year}{1991}), p.
  \bibinfo{pages}{156}.

\bibitem[{\citenamefont{Hukushima and Nemoto}(1996)}]{ptmc3}
\bibinfo{author}{\bibfnamefont{K.}~\bibnamefont{Hukushima}} \bibnamefont{and}
  \bibinfo{author}{\bibfnamefont{K.}~\bibnamefont{Nemoto}},
  \emph{\bibinfo{title}{Exchange {M}onte {C}arlo method and application to spin
  glass simulations}}, \bibinfo{journal}{J. Phys. Soc. Jpn.}
  \textbf{\bibinfo{volume}{65}}, \bibinfo{pages}{1604} (\bibinfo{year}{1996}).

\bibitem[{\citenamefont{Hukushima and Iba}(2003)}]{F}
\bibinfo{author}{\bibfnamefont{K.}~\bibnamefont{Hukushima}} \bibnamefont{and}
  \bibinfo{author}{\bibfnamefont{Y.}~\bibnamefont{Iba}}, in
  \emph{\bibinfo{booktitle}{The Monte Carlo Method In The Physical Sciences:
  Celebrating the 50th Anniversary of the Metropolis Algorithm}}, edited by
  \bibinfo{editor}{\bibfnamefont{J.~E.} \bibnamefont{Gubernatis}}
  (\bibinfo{publisher}{AIP}, \bibinfo{year}{2003}), vol. \bibinfo{volume}{690},
  pp. \bibinfo{pages}{200--206}.

\bibitem[{\citenamefont{Machta}(2010)}]{A}
\bibinfo{author}{\bibfnamefont{J.}~\bibnamefont{Machta}},
  \emph{\bibinfo{title}{Population annealing with weighted averages: A {M}onte
  {C}arlo method for rough free-energy landscapes}}, \bibinfo{journal}{Phys.
  Rev. E} \textbf{\bibinfo{volume}{82}}, \bibinfo{pages}{026704}
  (\bibinfo{year}{2010}).

\bibitem[{\citenamefont{Machta and Ellis}(2011)}]{B}
\bibinfo{author}{\bibfnamefont{J.}~\bibnamefont{Machta}} \bibnamefont{and}
  \bibinfo{author}{\bibfnamefont{R.}~\bibnamefont{Ellis}},
  \emph{\bibinfo{title}{{M}onte {C}arlo methods for rough free energy
  landscapes: Population annealing and parallel tempering}},
  \bibinfo{journal}{J. Stat. Phys.} \textbf{\bibinfo{volume}{144}},
  \bibinfo{pages}{541} (\bibinfo{year}{2011}).

\bibitem[{\citenamefont{Zhou and Chen}(2010)}]{ZhCh10}
\bibinfo{author}{\bibfnamefont{E.}~\bibnamefont{Zhou}} \bibnamefont{and}
  \bibinfo{author}{\bibfnamefont{X.}~\bibnamefont{Chen}}, in
  \emph{\bibinfo{booktitle}{Proceedings of the 2010 Winter Simulation
  Conference (WSC)}} (\bibinfo{year}{2010}), pp. \bibinfo{pages}{1211--1222}.

\bibitem[{\citenamefont{Wang et~al.}(2014)\citenamefont{Wang, Machta, and
  Katzgraber}}]{WaMaKa14}
\bibinfo{author}{\bibfnamefont{W.}~\bibnamefont{Wang}},
  \bibinfo{author}{\bibfnamefont{J.}~\bibnamefont{Machta}}, \bibnamefont{and}
  \bibinfo{author}{\bibfnamefont{H.~G.} \bibnamefont{Katzgraber}},
  \emph{\bibinfo{title}{Evidence against a mean-field description of
  short-range spin glasses revealed through thermal boundary conditions}},
  \bibinfo{journal}{Phys. Rev. B} \textbf{\bibinfo{volume}{90}},
  \bibinfo{pages}{184412} (\bibinfo{year}{2014}).

\bibitem[{\citenamefont{Doucet et~al.}(2001)\citenamefont{Doucet, de~Freitas,
  and Gordon}}]{DoFrGo01}
\bibinfo{editor}{\bibfnamefont{A.}~\bibnamefont{Doucet}},
  \bibinfo{editor}{\bibfnamefont{N.}~\bibnamefont{de~Freitas}},
  \bibnamefont{and} \bibinfo{editor}{\bibfnamefont{N.}~\bibnamefont{Gordon}},
  eds., \emph{\bibinfo{title}{Sequential {M}onte {C}arlo Methods in Practice}}
  (\bibinfo{publisher}{Springer-Verlag, New York}, \bibinfo{year}{2001}).

\bibitem[{\citenamefont{Rom{\'a} et~al.}(2009)\citenamefont{Rom{\'a},
  Risau-Gusman, Ramirez-Pastor, Nieto, and Vogel}}]{RoRiRaNiVo09}
\bibinfo{author}{\bibfnamefont{F.}~\bibnamefont{Rom{\'a}}},
  \bibinfo{author}{\bibfnamefont{S.}~\bibnamefont{Risau-Gusman}},
  \bibinfo{author}{\bibfnamefont{A.~J.} \bibnamefont{Ramirez-Pastor}},
  \bibinfo{author}{\bibfnamefont{F.}~\bibnamefont{Nieto}}, \bibnamefont{and}
  \bibinfo{author}{\bibfnamefont{E.~E.} \bibnamefont{Vogel}},
  \emph{\bibinfo{title}{The ground state energy of the {E}dwards-{A}nderson
  spin glass model with a parallel tempering {M}onte {C}arlo algorithm}},
  \bibinfo{journal}{Physica A: Statistical Mechanics and its Applications}
  \textbf{\bibinfo{volume}{388}}, \bibinfo{pages}{2821} (\bibinfo{year}{2009}).

\bibitem[{\citenamefont{Moreno et~al.}(2003)\citenamefont{Moreno, Katzgraber,
  and Hartmann}}]{formula}
\bibinfo{author}{\bibfnamefont{J.~J.} \bibnamefont{Moreno}},
  \bibinfo{author}{\bibfnamefont{H.~G.} \bibnamefont{Katzgraber}},
  \bibnamefont{and} \bibinfo{author}{\bibfnamefont{A.~K.}
  \bibnamefont{Hartmann}}, \emph{\bibinfo{title}{{Finding low-temperature
  states with parallel tempering, simulated annealing and simple {m}onte
  {c}arlo}}}, \bibinfo{journal}{International Journal of Modern Physics C}
  \textbf{\bibinfo{volume}{14}}, \bibinfo{pages}{285 } (\bibinfo{year}{2003}).

\bibitem[{\citenamefont{Douc and O.Capp\'e}(2005)}]{math}
\bibinfo{author}{\bibfnamefont{R.}~\bibnamefont{Douc}} \bibnamefont{and}
  \bibinfo{author}{\bibnamefont{O.Capp\'e}}, in
  \emph{\bibinfo{booktitle}{Proceedings of the 4th International Symposium on
  Image and Signal Processing and Analysis (ISPA)}} (\bibinfo{publisher}{IEEE},
  \bibinfo{year}{2005}), pp. \bibinfo{pages}{64--69}.

\bibitem[{\citenamefont{Wang et~al.}((2014), in preparation)\citenamefont{Wang,
  Machta, and Katzgraber}}]{pamc}
\bibinfo{author}{\bibfnamefont{W.}~\bibnamefont{Wang}},
  \bibinfo{author}{\bibfnamefont{J.}~\bibnamefont{Machta}}, \bibnamefont{and}
  \bibinfo{author}{\bibfnamefont{H.~G.} \bibnamefont{Katzgraber}},
  \emph{\bibinfo{title}{{Population annealing for large scale spin glass
  simulations}}} (\bibinfo{year}{(2014), in preparation}).

\bibitem[{\citenamefont{Yucesoy}(2013)}]{burcu}
\bibinfo{author}{\bibfnamefont{B.}~\bibnamefont{Yucesoy}}, Ph.D. thesis,
  \bibinfo{school}{University of Massachusetts Amherst} (\bibinfo{year}{2013}).

\bibitem[{sgs()}]{sgserver}
\emph{\bibinfo{title}{University of {C}ologne spin glass server}},
  \bibinfo{howpublished}{\url{http://www.informatik.uni-koeln.de/spinglass/}}.

\bibitem[{\citenamefont{Palassini and Young}(1999)}]{gse}
\bibinfo{author}{\bibfnamefont{M.}~\bibnamefont{Palassini}} \bibnamefont{and}
  \bibinfo{author}{\bibfnamefont{A.~P.} \bibnamefont{Young}},
  \emph{\bibinfo{title}{Triviality of the ground state structure in {I}sing
  spin glasses}}, \bibinfo{journal}{Phys. Rev. Lett.}
  \textbf{\bibinfo{volume}{83}}, \bibinfo{pages}{5126 } (\bibinfo{year}{1999}).

\end{thebibliography}

\end{document}